# Outcome prediction and individualized treatment effect estimation in patients with large vessel occlusion stroke

Lisa Herzog[1], Pascal Bühler[2], Ezequiel de la Rosa[3] Beate Sick[2,4], Susanne Wegener[1]

[1]Department of Neurology, University and University Hospital Zurich, Zurich, Switzerland
[2]Institute of Data Analysis and Process Design, Zurich University of Applied Sciences, Winterthur, Switzerland
[3]Department of Quantitative Biomedicine, University of Zurich, Zurich, Switzerland
[4]Epidemiology, Biostatistics and Prevention Institute, University of Zurich, Zurich, Switzerland
`lisa.herzog@uzh.ch`

**Abstract.** Mechanical thrombectomy has become the standard of care in patients with stroke due to large vessel occlusion (LVO). However, only 50% of successfully treated patients show a favorable outcome. We developed and evaluated interpretable deep learning models to predict functional outcomes in terms of the modified Rankin Scale score alongside individualized treatment effects (ITEs) using data of 449 LVO stroke patients from a randomized clinical trial. Besides clinical variables, we considered non-contrast CT (NCCT) and angiography (CTA) scans which were integrated using novel foundation models to make use of advanced imaging information. Clinical variables had a good predictive power for binary functional outcome prediction (AUC of 0.719 [0.666, 0.774]) which could slightly be improved when adding CTA imaging (AUC of 0.737 [0.687, 0.795]). Adding NCCT scans or a combination of NCCT and CTA scans to clinical features yielded no improvement. The most important clinical predictor for functional outcome was pre-stroke disability. While estimated ITEs were well calibrated to the average treatment effect, discriminatory ability was limited indicated by a C-for-Benefit statistic of around 0.55 in all models. In summary, the models allowed us to jointly integrate CT imaging and clinical features while achieving state-of-the-art prediction performance and ITE estimates. Yet, further research is needed to particularly improve ITE estimation.

**Keywords:** Interpretable Deep Learning, Functional Outcome Prediction, Individualized Treatment Effects.

## 1 Introduction

Stroke is a major cause of long-term disability and death world wide, with ischemic stroke accounting for more than 80% of all cases [1]. Among these, up to 46% suffer a large vessel occlusion (LVO) which affects one of the major cerebral arteries, like the internal carotid or middle-cerebral artery, and is associated with more severe neurological deficits and a doubled risk of long-term disability or death compared to other occlusion types [2]. In LVO stroke patients, early prognoses and optimal treatment decision-making remains particularly challenging.

Endovascular treatment such as mechanical thrombectomy (MT) has become the standard of care in LVO stroke; often used in combination with intravenous thrombolysis. Its efficacy was proven in multiple major randomized clinical trials (RCTs) across different patient groups including those with imaging confirmed large vessel occlusion [3], favorable core-penumbra profile [4], small [5, 4] and large core infarcts [6]. These RCTs highlight the critical role of imaging data in patient selection and outcome prediction which guides individualized risk assessment and treatment planning when combined with clinical features. Nonetheless, despite evidence-based patient selection and improvements in acute interventions, only approximately 50% of LVO stroke patients treated with MT show a favorable functional outcome [4, 7].

Predictive models incorporating clinical features and imaging modalities may enhance personalized decision-making. These models must estimate the potential functional outcome of MT versus no MT given a patient's individual characteristics. The individualized treatment effect (ITE) is the difference between predicted probabilities for a favorable functional outcome under the two treatment regimens. Unlike the average treatment effect (ATE), which summarizes treatment benefit averaged over all patients in the study population, the ITE considers heterogeneity in the treatment effect among patients by tailoring predicted probabilities to similar patient profiles [8].

Recent advances in machine and deep learning have improved functional outcome prediction in acute ischemic stroke using clinical data [9, 10] potentially combined with hand-crafted [11, 12] or raw imaging data [13, 14]. However, the majority of approaches relies on datasets including treated subjects only [11, 15, 13, 16] limiting their suitability for estimating counterfactuals. To the best of our knowledge, the MR PREDICTS [12] is the only tool estimating potential functional outcome under MT vs. no MT in LVO stroke, which, however, relies on clinical and extracted imaging features that are not available or unreliable at the time of decision making.

Here, we aimed at developing robust, interpretable and multi-modal models using raw non-contrast CT (NCCT) and angiography (CTA) scans alongside clinical features for predicting functional outcome in terms of the modified Rankin Scale (mRS) under MT vs. no MT to support treatment decision. The mRS is an ordinal scale ranging from 0 (no symptoms) to 6 (death). The models belongs to the class of ordinal neural network transformation models (ONTRAMs) which jointly integrate clinical and imaging data while preserving interpretability through odds ratio estimates quantifying the effect of clinical variables on the ordinal outcome. The models are developed on RCT data of the MR CLEAN trial and evaluated in a five-fold cross validation in terms of discrimination and calibration regarding prediction performance and ITE estimation.

## 2 Methods

### 2.1 Data

**Dataset** The MR CLEAN trial included 500 patients with imaging-confirmed LVO. Patients were randomized to either standard therapy (n=267) or endovascular treatment such as MT (n=233). No advanced neurovascular imaging profiles were used for inclusion. A detailed description of the cohort is provided in the original publication [3]. We selected the most relevant pre-treatment clinical data for model development based on expert-knowledge including patient characteristics (age), on admission scores (pre-stroke mRS, national institute of health stroke scale (NIHSS)), laboratory measures (glucose, systolic blood pressure), risk factors (diabetes, hypertension, smoking, previous ischemic event) and treatment related information (intravenous treatment (IVT), intraarterial treatment (MT), duration onset to door). In addition, non-contrast CT (NCCT) and CT angiography (CTA) scans collected before treatment were considered for model improvement.

**Data preparation** All models were implemented in a five-fold cross validation, i.e. data were split stratified into five comparable test sets regarding the outcome measure. Remaining data in the respective fold were used for training (85%) and validation (15%). Missing clinical data were imputed using k-nearest neighbor imputation. To ensure comparability between resulting parameter estimates, continuous variables were normalized and categorical variables dummy-encoded. Parameters for data preprocessing were estimated using the training data of the respective fold only.

To remove variation due to different data sources, NCCT images were resampled to an isotropic voxel size of 1x1x1mm$^3$. CTA volumes were cropped around the head to remove the neck and thorax, and were rigidly registered to the NCCTs using Elastix [17]. After skull stripping the registered volumes, the volumes were center-cropped to a target shape of 128x192x192. To analyze a combination of NCCT and CTA scans, the scans were concatenated along the first dimension after reducing the volumes to a size of 64x192x192 using average pooling. To focus on the most relevant brain tissue, Hounsfield Units were clipped to 0-80 and 50-400 for NCCT and CTA volumes, respectively.

### 2.2 Model development

**Ordinal neural network transformation models** The developed models belong to the class of ONTRAMs [18]. ONTRAMs are interpretable, modular approaches estimating probabilities for K classes of an ordinal outcome $Y$ using a distribution function $F_Y(y_k|x, I)$ based on multi-modal input data such as clinical features $x$ and images $I$ (see Fig. 1). All input modalities are modelled with (potentially deep) neural networks (NNs) whose outputs enter a so-called transformation function $h(y_k|x, I)$ – here $\vartheta_k - x^T\beta - \eta(I)$ – mapping a latent, pre-defined distribution $F_Z$ to the distribution of interest, such that

$$F_Y(y_k|x, I) = F_Z(\vartheta_k - x^T\beta - \eta(I)).$$

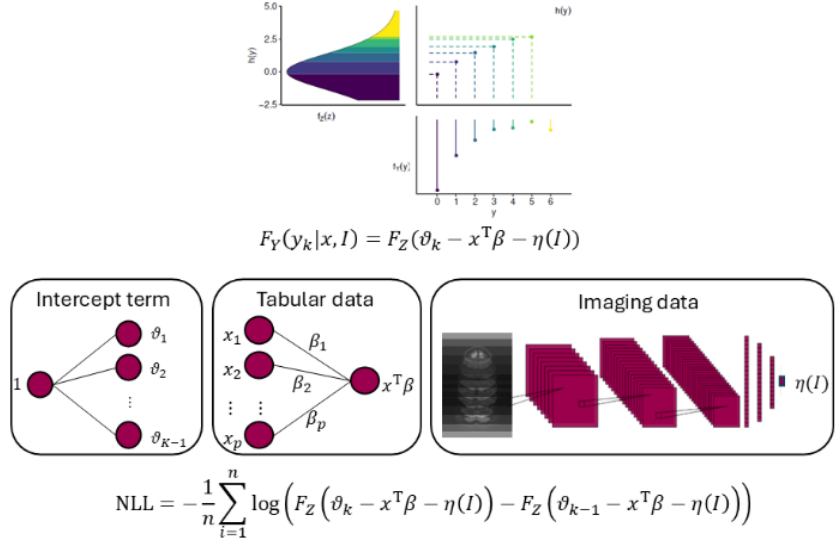

**Fig. 1.** The figure summarizes the ONTRAM framework when including clinical features $x$ as linear predictors and imaging data $I$ with an advanced NN architecture. The probabilities of the outcome distribution $f_Y(y_k|x, I)$ are estimated via a transformation function $h$ from the area under the curve over the respective bins of the standard logistic distribution $f_Z(z)$. The terms of $h$ are controlled with NNs which are jointly trained by minimizing the NLL.

That is, instead of estimating the outcome distribution directly, we specify $F_Z$ and estimate the parameters of the transformation function. The NNs are jointly fitted by minimizing the negative log-likelihood

$$\text{NLL} = -\frac{1}{n}\sum_{i=1}^{n} log\left(F_Z(\vartheta_k - x^T\beta - \eta(I)) - F_Z(\vartheta_{k-1} - x^T\beta - \eta(I))\right),$$

using standard optimization algorithms like Adam. Implementations of the ONTRAM framework are available on Github[1].

Here, the baseline transformation function is modelled with $\vartheta_k, k = 1, ..., 6$. Clinical variables enter the function as linear predictors $x^T\beta$ which shifts $\vartheta_k$'s up- and downward to change the outcome distribution accordingly. $F_Z$ is defined as standard logistic distribution which enables to interpret the parameters $\beta$ as log-odds ratios, meaning that if $x_p$ is increased by one unit and all other predictors stay constant, then the predicted odds to belong to a higher class than $y_k$ is changing by $e^{\beta_k}$. The clinical variables include a treatment indicator which is set to 1 and 0 for estimating the ordinal outcome distribution under MT vs. no MT, respectively. Summing up the estimated probabilities for mRS 0-2 yields predicted probabilities for a favorable functional outcome under MT vs. no MT that are subtracted to estimate the ITE. This approach is assumed to yield causally interpretable ITEs given that identifiability assumptions are in place as in case of a valid RCT [8].

---

[1] See https://github.com/liherz/ontram_pytorch for an implementation in Pytorch.

In this cohort, no significant interactions between treatment and clinical variables were found [3]. As a result, no terms for treatment-covariate interactions were incorporated, i.e. a homogenous treatment effect was assumed. Please note, this still leads to variability in ITEs due to the non-linearity of $F_Z$. The reason for this variability is, however, not a differential effect of treatment but variability in clinical and imaging data [8].

To make use of the most relevant imaging information, NCCT and CTA scans were included into ONTRAMs using pre-trained encoders of novel foundation models such as a (1) Swin UNETR, a vision transformer trained on 5,050 publicly available CT scans from several different body organs [19] and a (2) 3D convolutional neural network based on a SegResNet architecture trained on 148.000 CT scans [20]. To each encoder, we attached a classification head with ReLU activation that consisted of a 3D average pooling layer followed by two fully-connected layers with 256 and 128 features. The output layer consisted of one unit which is the input $\eta(I)$ to the transformation function. For regularization, dropout of 0.3 was introduced into the classification head.

**Implementation and training procedure** We first implemented an ONTRAM based on clinical data only. This proportional odds model was trained for 10.000 epochs to ensure convergence using a batch size of 128 and the default parameters of the Adam optimizer. The NNs for estimating the intercepts $\vartheta_k$ and tabular data $x^T\beta$ in ONTRAMs additionally relying on imaging data are initialized with these pre-trained weights (compare Fig. 1). To train the imaging data part, we fix the weights of the pre-trained encoder and train the classification head with a batch size of 2, learning rate of 1e-4 and weight decay of 1e-6. Within 150 epochs, the whole network is fine-tuned with a learning rate of 1e-6. Performance dropped significantly when tabular and imaging data were trained simultaneously (not shown here).

Data augmentation was used to increase variability during model training including random spatial cropping to a size of 128x128x128, rotation and shifting, while noise, contrast and brightness were randomly adapted. HU values were normalized to the interval of [0,1] before feeding the data into the foundation models. At test time, center spatial cropping and normalization was used to have comparable input data for model evaluation.

### 2.3 Evaluation

**Descriptive statistics** To summarize the patient data used for outcome prediction, descriptive statistics were calculated. Categorical variables are described using frequencies and percentages. Continuous variables are summarized with medians and interquartile ranges. Measures are reported for all patients and those with favorable (mRS of 0-2) vs. unfavorable (mRS of 3-6) functional outcome.

**Metrics** All models were evaluated on the test data in terms of discrimination and calibration with respect to binary functional outcome prediction (mRS 0-2 vs. 3-6). Therefore, predicted probabilities for mRS 0-2 and 3-6 were summed up. Binary functional outcome prediction is more relevant for clinical application, however, in previous work, we could show that it is advantageous to make use of the full information during model training, which is why all models were trained on the ordinal outcome [13]. Besides the area under the receiver operator characteristics curve (AUC) highlighting discriminatory ability, NLL and Brier scores are summarized to consider calibration. For evaluating model calibration for ITE estimation, we compare the ATE to the average across ITEs. In addition, we summarized the C-for-Benefit statistic [21] measuring discrimination ability, i.e. if patients predicted to benefit more from treatment, actually do. Metrics are computed for 1,000 bootstrap samples to derive 95% confidence intervals with the basic bootstrap method.

## 3   Results

Of the 500 patients in the MR CLEAN trial, 449 were considered for analysis. For the remaining patients, CT imaging was incomplete and/or could not be registered. The estimated ATE in terms of the odds ratio for favorable vs. unfavorable functional outcome was 2.09 (2.05 in the original cohort). The percentage of patients with a favorable functional outcome increased from 19.25% in the control to 33.33% in the treatment group. Patients with favorable functional outcome were younger, had lower glucose, systolic blood pressure and NIHSS values, less often suffered diabetes and hypertension and were more often smoking cigarettes (s. Table 2).

Table 1 summarizes model performance in terms of outcome prediction and ITE estimation when considering different input data and one of two foundation models for integrating CT imaging. Both foundation models yielded comparable results which is why the following results refer to the models based on the pre-trained SwinUNETR encoder. The best prediction performance was achieved in the model using clinical and CTA imaging with an AUC of 0.737 [0.687, 0.795] and a Brier Score of 0.168 [0.148, 0.187]. However, the performance gain compared to clinical data alone was non-significant (AUC of 0.719 [0.666, 0.774]). In line with this, adding NCCT or a combination of NCCT and CTA to clinical data did not improve prediction performance (AUC of 0.717 [0.666, 0.774]) as indicated by heavily overlapping confidence intervals. Regarding ITE estimation, all models performed similarly, with C-for-Benefit values around 0.55. Again, adding CTA imaging to clinical data showed a small but nonsignificant improvement in prediction performance. The difference between the ATE and mean across ITEs is close to zero indicating that estimated ITEs are well-calibrated to the observed ATE. However, on average all models slightly underestimated the ATE. As in case of prediction performance, clinical features seemed to carry most of the predictive signal for ITE estimation.

**Table 2.** Frequencies (percentages) and medians [interquartile ranges] for patients with favorable (mRS 0-2) and unfavorable (mRS 3-6) functional outcome.

|  | Unfavorable (n=333) | Favorable (n=116) | P-value |
|---|---|---|---|
| IAT (yes) | 140 (42.0) | 70 (60.3) | 0.001 |
| IVT (yes) | 292 (87.7) | 107 (92.2) | 0.241 |
| Age (in years) | 67 [57, 77] | 62 [49, 72] | <0.001 |
| Glucose (mmol/l) | 6.80 [5.90, 8.10] | 6.25 [5.77, 7.12] | 0.004 |
| Pre-mRS 0-2 | 76 (22.8) | 14 (12.1) | 0.018 |
| NIHSS | 18 [15, 22] | 15 [11, 19] | <0.001 |
| Syst. BP (mmHg) | 146 [130, 161] | 137 [125, 150] | <0.001 |
| Diabetes (yes) | 52 (15.6) | 8 (6.9) | 0.027 |
| Hypertension (yes) | 161 (48.3) | 42 (36.2) | 0.031 |
| Smoker (yes) | 82 (24.6) | 44 (37.9) | 0.009 |
| Prev. ischemic stroke (yes) | 42 (12.6) | 8 (6.9) | 0.130 |
| Onset to door (min.) | 98 [50, 197] | 90 [45, 189] | 0.336 |

**Table 1.** Results of binary functional outcome and ITE prediction when using different inputs and encoders for CT imaging (upper row: SwinUNETR, lower row 3D CNNSegResNet).

|  | Clinical (C) | C + NCCT | C + CTA | C + NCCT+CTA |
|---|---|---|---|---|
| NLL | 0.527 [0.47, 0.59] | 0.530 [0.47, 0.59] | **0.513 [0.46, 0.57]** | 0.527 [0.47, 0.58] |
|  |  | 0.539 [0.48, 0.6] | 0.530 [0.47, 0.59] | 0.530 [0.47, 0.59] |
| AUC | 0.719 [0.67, 0.77] | 0.717 [0.67, 0.77] | **0.737 [0.69, 0.8]** | 0.717 [0.67, 0.77] |
|  |  | 0.710 [0.66, 0.76] | 0.716 [0.66, 0.77] | 0.717 [0.67, 0.78] |
| Brier Score | 0.172 [0.15, 0.19] | 0.175 [0.16, 0.2] | **0.168 [0.15, 0.19]** | 0.173 [0.15, 0.19] |
|  |  | 0.176 [0.15, 0.12] | 0.174 [0.15, 0.19] | 0.173 [0.15, 0.19] |
| C-ben | 0.551 [0.48, 0.62] | 0.551 [0.49, 0.62] | 0.558 [0.49, 0.62] | 0.552 [0.49, 0.62] |
|  |  | **0.570 [0.5, 0.64]** | 0.560 [0.49, 0.63] | 0.549 [0.48, 0.62] |
| ATE-ØITE | **0.065 [-0.01, 0.15]** | 0.069 [-0.01, 0.15] | 0.068 [-0.02, 0.14] | 0.071 [-0.01, 0.16] |
|  |  | 0.068 [-0.01, 0.15] | 0.067 [-0.02, 0.15] | 0.070 [-0.02, 0.14] |

The most important predictor for functional outcome was functional disability before stroke followed by previous ischemic stroke, NIHSS on admission and treatment related information which is in line with the literature (see Fig. 2). The results were similar across all models.

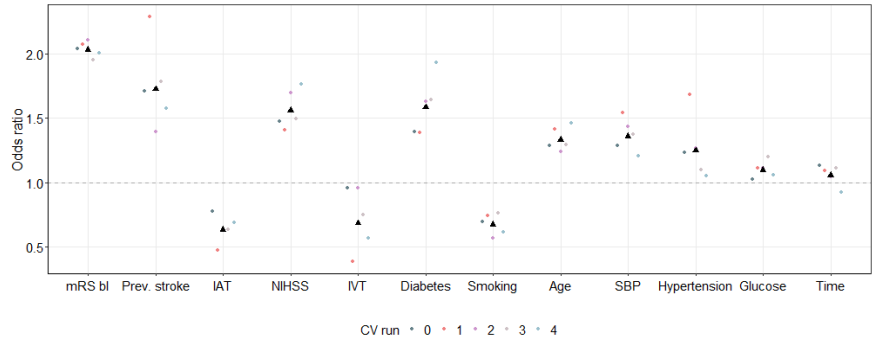

**Fig. 2.** Estimated odds ratios for clinical features when used in combination with CTA imaging to predict ordinal functional outcome. ORs>1 indicate a higher chance for a higher mRS class with respect to the reference/a one unit increase in standard deviation of the respective variable.

## 4      Discussion

In this work we developed and compared different deep learning models for treatment decision making by predicting potential functional outcome under MT vs. no MT using clinical features and CT imaging from a RCT. The models are integrated into the framework of ONTRAMs which jointly model multi-modal inputs while preserving interpretability for a part of the data. We integrated NCCT and CTA scans with novel foundation models for CT imaging and reached AUC prediction performances around 0.7 when additionally including clinical data. The best performing model achieved an AUC of 0.737 [0.687, 0.795] based on clinical and CTA imaging. In terms of treatment effect estimation, the models were well calibrated with respect to the ATE but had a low discriminatory ability highlighted by C-for-benefit statistic values around 0.55.

As in other studies, only slight improvement in prediction performance was found when adding CT imaging to clinical data indicated by heavily overlapping confidence intervals [22, 23, 24]. Here, we could confirm this observation regarding ITE estimation. In this work, we integrated a homogenous treatment effect because no significant interactions were identified in the original publication [3]. Nonetheless, in future projects, we additionally aim at integrating interactions for considering differential treatment effects and non-linear terms for specific variables like age. This was shown to improve functional outcome prediction slightly [12]. In ongoing projects, we additionally consider T- and X-learners which are trained on treated and untreated subjects for improved ITE estimation.

A limitation of our study is the small sample size. We accounted for this by using a cross-validation setting for analysis, considered pre-trained NNs for integrating CT imaging and performed data augmentation. Although we believe that our results are reliable, we also expect to improve performance with more data which will be added in future experiments. Nonetheless, our results are comparable to other studies which mainly consider all ischemic stroke cases and not the challenging subgroup of LVO. [25] achieved an AUC of 0.65 – 0.72 in ischemic stroke patients when analyzing clinical features with different machine learning approaches. [22] showed an AUC of 0.806 +/- 0.082 based on CTA and clinical data when considering 743 patients with ischemic strokes. In a subgroup of LVO stroke, [23] achieved AUCs between 0.6 and 0.77 when using clinical features and lesion loads of registered CT brain atlas registered to follow-up NCCT scans. The highest AUC of 0.85 (0.75-0.94) was achieved by [14] in LVO stroke when using NCCT and clinical data. However, they only used one split of training and test data – such a high AUC was also achieved in one split in our cross validation but not when considering all test patients (not shown here).

In summary, the proposed models provide a novel, interpretable deep learning framework for functional outcome prediction while enabling to estimate a treatment effect like in statistical models when using clinical features and raw CT imaging. Yet, further research is needed to better understand why models do not exceed AUC values of 0.8 and to improve estimates of treatment effect and patient outcome after LVO.